\documentclass[conference]{IEEEtran}
\IEEEoverridecommandlockouts
\usepackage{cite}
\usepackage{amsmath,amssymb,amsfonts}
\usepackage{algorithmic}
\usepackage{graphicx}
\usepackage{textcomp}
\usepackage{makecell}
\usepackage{xcolor}
\usepackage{wrapfig}
\usepackage{listings}
\usepackage[format=plain, font=small, belowskip=-15pt,aboveskip=2pt]{caption}

\def\BibTeX{{\rm B\kern-.05em{\sc i\kern-.025em b}\kern-.08em
    T\kern-.1667em\lower.7ex\hbox{E}\kern-.125emX}}
\begin{document}

\title{It's all about data movement: Optimising FPGA data access to boost performance}

\author{\IEEEauthorblockN{Nick Brown}
\IEEEauthorblockA{\textit{EPCC, The University of Edinburgh}\\
Bayes Centre, 47 Potterrow, Edinburgh, UK \\
n.brown@epcc.ed.ac.uk}
\and
\IEEEauthorblockN{David Dolman}
\IEEEauthorblockA{\textit{Alpha Data Parallel Systems}\\
160 Dundee Street, Edinburgh, UK}
}

\maketitle

\begin{abstract}
The use of reconfigurable computing, and FPGAs in particular, to accelerate computational kernels has the potential to be of great benefit to scientific codes and the HPC community in general. However, whilst recent advanced in FPGA tooling have made the physical act of programming reconfigurable architectures much more accessible, in order to gain good performance the entire algorithm must be rethought and recast in a dataflow style. Reducing the cost of data movement for all computing devices is critically important, and in this paper we explore the most appropriate techniques for FPGAs. We do this by describing the optimisation of an existing FPGA implementation of an atmospheric model's advection scheme. By taking an FPGA code that was over four times slower than running on the CPU, mainly due to data movement overhead, we describe the profiling and optimisation strategies adopted to significantly reduce the runtime and bring the performance of our FPGA kernels to a much more practical level for real-world use. The result of this work is a set of techniques, steps, and lessons learnt that we have found significantly improves the performance of FPGA based HPC codes and that others can adopt in their own codes to achieve similar results.
\end{abstract}



\section{Introduction}
The anticipated end of Moore's law has meant that interest in adopting novel computing techniques is receiving more and more attention, and one such area is that of Field Programmable Gate Arrays (FPGAs) which provide a large number of configurable logic blocks sitting within a sea of configurable interconnect. Based on recent advances in programming tooling developed by vendors, it is easier than ever before for developers to convert their algorithms down to a level which can configure these fundamental components and as-such execute their HPC codes in hardware rather than software. With the addition of other facets on the chip, such as fast block RAM (BRAM), Digital Signal Processing (DSP) slices, and high bandwidth connections off-chip, FPGAs are hugely versatile. Whilst this reconfigurable technology has a long heritage in embedded systems and signal processing, its adoption in scientific computing has, until now, been more limited. There are a number of reasons for this, but the combination of recent advances in high level programming tools and predicted end of CPU performance scaling means that the role that reconfigurable architectures can play in HPC is worth examining in detail. 

The Met Office NERC Cloud model (MONC) \cite{easc} is an open source high resolution modelling framework that employs Large Eddy Simulation (LES) to study the physics of turbulent flows and further develop and test physical parametrisations and assumptions used in numerical weather and climate prediction. As a major atmospheric model used by UK weather and climate communities, MONC replaces an existing model called the Large Eddy Model (LEM) \cite{lem} which was an instrumental tool, used by scientists, since the 1980s for activities such as development and testing of the Met Office Unified Model (UM) boundary layer scheme \cite{lock1998}, convection scheme \cite{petch2001} and cloud microphysics \cite{hill2014}. Scientists are continually demanding the ability to simulate larger domains, at increased accuracy, and at reduced time to solution. Therefore any opportunity to accelerate the model is important to explore and exploit. 

In previous work \cite{fpga} we ported the MONC advection kernel to FPGAs via High Level Synthesis (HLS) and demonstrated that it is not enough to simply copy code over from the CPU to the FPGA tooling and synthesise it. Instead, a programmer must change the entire way in which they approach their algorithms, moving to a much more \emph{dataflow} style \cite{ma2017optimizing} in order to achieve acceptable performance. Whilst we demonstrated around a one hundred time performance difference between un-optimised code and HLS kernels optimised for the FPGA, our previous FPGA based code was still significantly slower than running on the CPU. Using this previous work as a foundation, in this paper we describe further work where we developed an approach which enables us to first profile the code, highlighting exactly where the bottlenecks were occurring, and then significantly improve the performance of our FPGA kernel based on this information. 

Optimising data movement is critically important for all computing devices, and different techniques suit different technologies. To run HPC codes optimally on FPGAs we need to understand which approaches most effectively reduce the cost of data movement, and this is the central focus of our work here. The organisation of this paper is as follows, in Section \ref{sec:bg} we introduce the general background to this work as well as describe the previous work done porting the MONC advection kernel to FPGAs in detail. In Section \ref{sec:prof} we then explore our approach to profiling the HLS kernel and, based upon this information, the steps we took to further optimise and significantly reduce the runtime. Section \ref{sec:DMA} describes the work done to ameliorate DMA data transfer time to and from the PCIe board by overlapping compute with transfer, before we contrast the performance of our advection scheme on the FPGA against three common CPU microarchitectures popular in HPC machines in Section \ref{sec:performance}. Lastly, we draw conclusion and discuss further work in Section \ref{sec:conc}. 

\section{Background and related work}
\label{sec:bg}
There has been much development in programming tools for FPGAs and the use of higher level programming abstractions such as High Level Synthesis (HLS) is amongst the most prevalent of these. In HLS a kernel is written in C, C++ or System C and then automatically translated into the underlying Hardware Description Language (HDL), with the programmer not needing to express their code at the very low level of HDL. Within their high level code programmers are able to direct the tooling via pragma style hints, and this HLS approach substantially speeds up development time. However, HLS is not a silver bullet and whilst the physical act of programming FPGAs has become much easier, one must still \emph{think dataflow} to get good performance.

In this work we follow the high-level productivity design methodology \cite{highproduct}, where one explicitly writes a C kernel in HLS, generates the HDL and exports this as an IP block. The FPGA is then configured using a block design approach, where existing IP blocks including the programmer's HLS kernel IP block(s), are imported and connected together. At the block design level one is combining their kernel IP block with other infrastructure necessary for running their code, such as DRAM controllers, and we followed this approach because we felt it gave us more control over the configuration of our design. We also believe that lessons learnt for HPC codes can feed into higher level abstractions, such as those employed by OpenCL \cite{opencl} and SDAccel \cite{sdaccel}.

There have been a number of previous activities investigating the role that FPGAs can play in accelerating HPC codes. One such example is \cite{lfricfpga}, where the authors investigated using the high-level productivity design methodology via HLS to accelerate solving the Helmholtz equation. They offloaded the matrix-vector updates which are required as part of this solver onto a Zynq Ultrascale, however the performance obtained was around half of that when the code was run on a twelve core Broadwell CPU. In \cite{lfricfpga} the author's matrix-vector kernel involved looping over two double precision floating point operations, whereas in this work we are focused on accelerating a much more complicated kernel comprising of fifty three double precision floating point operations per grid cell. These double precision operations involve twenty one additions or subtractions, and thirty two double precision multiplications. In \cite{lfricfpga} the authors were limited to a maximum data size of 17MB due to BRAM limits on the Zynq, whereas the work detailed in this paper contains experiments with grid sizes of up to 6.44GB of data (and a further 6.44GB for resulting values), necessitating the use of external DRAM on the PCIe FPGA card.

\subsection{Hardware setup}
For the work described in this paper we are using an ADM-PCIe-8K5 PCI Express card, manufactured by Alpha Data, which mounts a Xilinx Kintex Ultrascale KU115-2 FPGA. This FPGA contains 663,360 LUTs, 5520 DSP48E slices and 4320 BRAM 18K blocks. The card also contains two banks of 8GB DDR4-2400 SDRAM, external to the FPGA, and a number of other interfaces which are not relevant to this work. The host CPU and board interact via a PCIe Gen3 * 8 interface. Because the FPGA used for this work is part of the Xilinx product family it is their general ecosystem, including tooling (version 2018.3), that we use in this work. However, we believe that the lessons learnt apply more generally to product families of FPGAs from other vendors too.

This PCIe card is plugged into an Intel Xeon system, which contains two Sandybridge CPUs, each with four physical cores running at 2.40GHz, and 32GB RAM (16GB per NUMA region). Whilst some FPGAs such as the Zynq use a more embedded style, where typically ARM cores are combined with FPGA fabric on the same chip, we believe this PCIe setup is more interesting in the field of HPC because a powerful CPU can be used on the host side and secondly because a large amount of memory can be placed close to the FPGA on the PCIe card to handle the acceleration of large problems. 

\subsection{The existing MONC advection FPGA kernel}

Advection is the movement of grid values through the atmosphere due to wind, and in \cite{fpga} we focused on porting MONC's Piacsek and Williams \cite{pwadvection} advection scheme onto a Xilinx Kintex Ultrascale KU115-2 FPGA using HLS. At  around 50\% of the overall runtime, this kernel is the single longest running piece of functionality and the code loops over three fields;  \emph{U}, \emph{V} and \emph{W}, representing wind velocity in the \emph{x}, \emph{y} and \emph{z} dimensions respectively. Called each timestep of the model, calculating the advection results, otherwise known as the \emph{source terms}, of the three fields for each grid cell involves 53 double precision operations. This advection kernel is a stencil based code, of depth one, requiring neighbouring values across all the three dimensions per grid cell.

\begin{figure}
\centering
\includegraphics[scale=0.3]{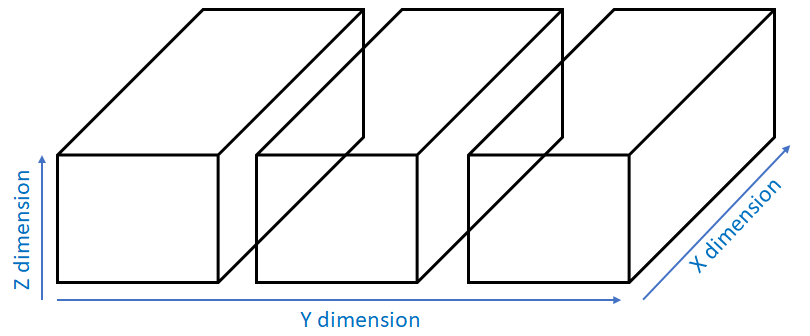}
\caption{Illustration of domain decomposition for a single HLS advection kernel of data into separate 3D blocks}
\label{fig:domaindecomp}
\end{figure}

In \cite{fpga} we focused on using HLS to optimise the computational part of the advection kernel. Because the code is stencil based, the naive memory access pattern is that of fetching all values needed for computing a grid cell from the on-board, but off-chip, DRAM. However, as we are moving through the points linearly this is wasteful, because much of the data will have already been fetched for preceding grid cells. Instead, we restructured the CPU code to work in 3D blocks of data as per Figure \ref{fig:domaindecomp} based on the fact that only the current, previous and next slices of data in the \emph{z} and \emph{y} dimensions are required to calculate each grid point in a current \emph{z} by \emph{y} slice. On-chip BRAM was used to hold these slices of data and when advancing to the next slice in the \emph{x} dimension, the kernel shifts the current slices of data down by one, discarding the current previous slice of data, and retrieves the \emph{x+1} slice. The size of each \emph{z} by \emph{y} slice is determined by the amount of BRAM available, and we use a size of 64 by 64. We found that this approach was well worth it as, with a pipelined code in the vertical \emph{z} dimension and with slice size of 64 by 64, it reduced the runtime by around three and a half times. 

\begin{lstlisting}[frame=lines,caption={Sketch of HLS code structure},label={lst:hls_kernel}, numbers=left, belowcaptionskip=-3pt, language=c]
for (unsigned int m=start_y;m<end_y;m+=BLOCKSIZE_IN_Y) {
  for (unsigned int c=0; c < slice_size; c++) {
    #pragma HLS PIPELINE II=1
    // Read in from DRAM to load up first two slices for block
    ...
  }

  for (unsigned int i=start_x;i<end_x;i++) {
    for (unsigned int c=0; c < slice_size; c++) {
      #pragma HLS PIPELINE II=1
      // Move data in slice+1 and slice down by one in X dimension
      ...
    }

    for (unsigned int c=0; c < slice_size; c++) {
      #pragma HLS PIPELINE II=1
      // Load data for U field from DRAM
      u_vals[c]=u[read_index];
      ...
    }
    
    // Also do the same load for V and W in separate loops
    ...

    for (unsigned int j=0;j<number_in_y;j++) {
      for (unsigned int k=1;k<size_in_z;k++) {
        #pragma HLS PIPELINE II=1
        // Do calculations for U, V, W field grid point 
		...
		
        su_vals[jk_index]=su_x+su_y+su_z;
        sv_vals[jk_index]=sv_x+sv_y+sv_z;
        sw_vals[jk_index]=sw_x+sw_y+sw_z;
      }
    }
		
    for (unsigned int c=0; c < slice_size; c++) {
      #pragma HLS PIPELINE II=1
      // Write data for SU field to DRAM
      su[write_index]=su_vals[c];
      ...
    }
    
    // Also do the same write for SV and SW in separate loops
    ...
  }
}
\end{lstlisting}

Listing \ref{lst:hls_kernel} sketches the structure of our HLS kernel code, looping over the separate 3D blocks at line 1. The kernel will, for each 3D block, initially load in the first two slices from DRAM for that block between lines 2 and 6, before looping over each slice in the \emph{x} dimension (line 8). As the code progresses from one slice to the next it will shift the locally stored slice data down by one (lines 9 to 13) and then retrieve data for the \emph{x+1} slice, illustrated for the \emph{U} field between lines 15 and 20. In our design all field array variables share the same physical HLS port and as such there can only be one access per clock cycle on any variable, so the loop at lines 15 to 20 is replicated for the \emph{V} and \emph{W} fields, which has been omitted for brevity. 

Once the required slice data has been retrieved this is then looped over, with the inner vertical loop in the \emph{z} dimension pipelined. It is this inner pipelined loop that contains the 53 double precision operations and result values are calculated which are then written into the corresponding locations of the \emph{su\_vals}, \emph{sv\_vals}, and \emph{sw\_vals} BRAM arrays, holding resulting source terms for the slice. Once the calculations for the current slice have completed, the local BRAM result data arrays are written out to off-chip DRAM, with an illustration for the \emph{su} field provided between lines 37 and 42. Whilst it may seem strange that the code contains explicit copies rather than calling out to the \emph{memcpy} function, we found that a pipelined loop to perform this copy resulting in slightly better performance \cite{fpga}.


Structuring the code as per Listing \ref{lst:hls_kernel}, and other optimisations such tuning the double precision HLS cores to support a clock frequency of 310 Mhz, and optimising the HLS data ports to tune for latency and run in burst mode, provided a 100 time performance difference between the initial HLS kernel unmodified from the CPU code, and the optimised version. The HLS kernel itself was around 25\% faster than a single core of Sandybridge and comparable to the performance provided by a single core of Broadwell. However when factoring in the DMA data transfer time, copying field data from the host to the PCIe board and resulting values back again, along with comparing an entire CPU with multiple cores against the FPGA with multiple advection cores, our approach performed around 7 times slower than an 18-core Broadwell for the largest problem size.


There were two main reasons for this poor performance. Whilst the performance of a single advection HLS kernel was comparable to a single core of Broadwell, we could only fit 12 advection kernels onto the FPGA against 18 cores of Broadwell. Furthermore, the reports from Vivado HLS suggested a significantly lower runtime than we were actually achieving, indicating significant overhead when the kernels were executed in practice. Secondly, our strategy for data transfer between the host and PCIe card was very simplistic, where the kernels would not start until all the data had been transferred, and data transfer of results back to the host would not start until all the kernels had completed. This added significantly to the runtime, especially with experiments run at larger problem sizes where DMA transfer time accounted for over 70\% of the overall runtime with a grid size of 267 million points.

\subsection{Advection kernel GPU port}
We have previously explored the acceleration of this advection kernel using GPUs via OpenACC \cite{brown2015directive}. Whilst this approach initially demonstrated some promise, the cost of data transfer to and from the GPU card via PCIe was the dominant factor, ultimately suffering from the same issues we faced in \cite{fpga}. The implementation relied on OpenACC's \emph{async} clause, exploiting CUDA streams \cite{luitjens2015cuda} which define a queue of work including both computational and data transfer tasks. The idea is that, because activities in the queue are predefined, then as one tasks completes the next automatically initiates without any involvement required from the CPU.

The streaming approach of queuing up work is very powerful, and has been a critical part in delivering good GPU performance in numerous codes \cite{papenhausen2013rapid}. However, we found in our previous work with MONC that when driven by OpenACC, exploitation of streams was fairly coarse-grained. Even though CUDA supports the definition of many streams, what dominated our GPU performance was the fact that we were unable to effectively split iterations of a single computational loop up into separate streams. We believe that this streaming approach is of great applicability to FPGAs too, but crucially one must be able to split the data transfer into chunks corresponding to independent iterations of a single computational kernel.

\section{Optimising the HLS kernel}
\label{sec:prof}
The latency and trip count of Vivado HLS's scheduling viewer, in combination with the clock frequency of the core, suggested that the performance achieved in practice was significantly lower than what it should be. We felt that much of this was overhead in accessing the on-card but off-chip DRAM. Whilst simulation can be used to gather timing information, this does not take into account these real-world factors that are crucial to achieving high performance. A major limitation of Vivado HLS is that it does not provide any ability to profile the code when it is actually physically executing on the FPGA. 

It was our hypothesis that the time transitioning from one 3D block to the next was very costly because the pipeline had to be entirely drained and two slices loaded from DRAM before any computation could proceed. Whilst it is possible to add traces onto specific AXI-4 connections via the debug interface in Vivado, and then view plots of these when these ports are busy or idle, the code in Listing \ref{lst:hls_kernel} accesses DRAM throughout its execution and this debugging information does not tell us where about in the code the bottlenecks lie. Without any further tools it was difficult to estimate exactly where the bottleneck(s) lie and as such we needed to gather more concrete numbers to really understand what was going on, and where optimisation effort should be focused.

\subsection{Profiling: if you cant measure it you can't improve it}

When programming in C with HLS it is not possible to explicitly gain access to the clock signal, and the way that HLS works means that functionality can not run concurrently as one expects on the CPU. As such we wrote a separate profiler IP block, using HLS, that could be connected to our advection HLS kernel via Vivado block design. The advection kernel contains multiple blocks of code, each of which needed to be timed and a running total for each maintained. Every time a block of interest is executed by the advection kernel, the profiler is informed at the start and end of execution, with the block's runtime added to a running total for the block. 

\begin{lstlisting}[frame=lines,caption={Illustration of interaction with the profiling IP block},label={lst:hls_profile}, belowcaptionskip=-3pt, language=c]
profiler_commands->write(BLOCK_1_START);
ap_wait();

function_to_execute(.....);

ap_wait();
{
  #pragma HLS protocol fixed
  profiler_commands->write(BLOCK_1_END);
  ap_wait();
}
\end{lstlisting}

The profiler is driven by commands from the HLS advection kernel, which are numeric values read from an \emph{hls stream} interface. This mechanism is used to not only start and stop the profiling of distinct blocks of code, but also to initialise the profiler and retrieve final timings for each block. Listing \ref{lst:hls_profile} sketches the interaction between the HLS advection kernel and profiler, illustrating a block of code to be timed (denoted by the \emph{function\_to\_execute} call, although this does not need to be a function call.) The \emph{profiler\_command} variable is an HLS stream that connects the advection kernel to the profiler IP block and commands are sent denoting the start or end of a block. The \emph{ap\_wait} call is used to delay for a clock cycle and the \emph{HLS protocol fixed} pragma informs HLS not to reorder calls in that block. These were required because HLS does not detect a dependency between the writing to \emph{profiler\_commands} and the execution of the function of interest. As such, it has a tendency to place the ending call before, or concurrent to, the function to profile. We found Vivado HLS's schedule viewer an important tool here, that enabled us to track down where these writes were being issued. Based on this information, constructs to control placement were added to ensure that the profiler was being started and stopped at the appropriate point in relation to the block of code being timed. 

\begin{figure}
\centering
\includegraphics[scale=0.45]{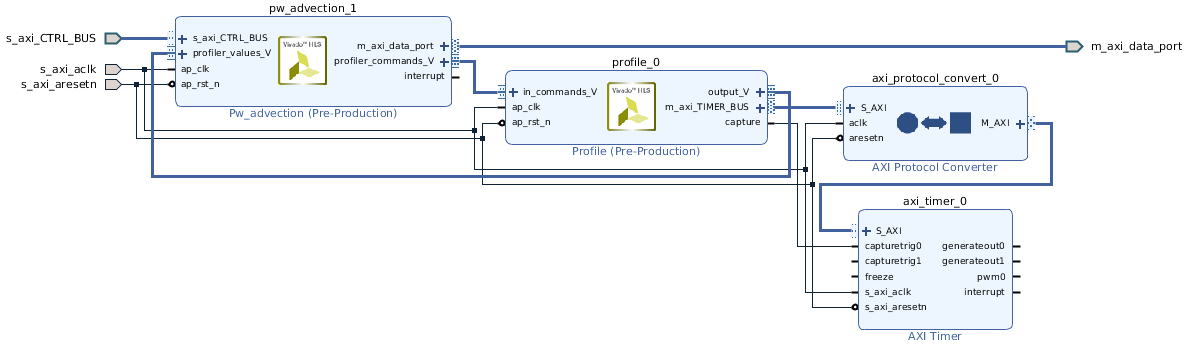}
\caption{Block design of PW advection IP connected to profiling IP block and AXI timer}
\label{fig:profiling}
\end{figure}

Figure \ref{fig:profiling} illustrates our advection hierarchical block in Vivado block design. The profiler IP block is connected to the advection kernel, and the profiler is also connection to Xilinx's \emph{AXI timer}. This timer increments a 64 bit value on every clock cycle and the timer is run in capture mode, which stores the counter value in a separate register when an interrupt is raised on the \emph{capture} port by the profiler IP block. It is this captured value that is then read by our profiling IP block via the AXI4-Lite port. Whilst this is a simple approach to profiling, it is beyond anything provided by Vivado HLS, and sufficient to generate simple timing numbers for different parts of our HLS kernel code to understand what is going on.

\subsection{Optimising the advection HLS kernel}
\label{sec:optimisinghls}

\begin{figure*}
 \centering
\begin{tabular}{ | c c c c c c | }
\hline
\textbf{Description} & \makecell{\textbf{Total runtime} \\ \textbf{(ms)}} & \makecell{\textbf{\% time spent} \\ \textbf{doing compute}} & \makecell{\textbf{Load data} \\ \textbf{(ms)}} & \makecell{\textbf{Prepare stencil} \\ \textbf{ \& compute results (ms)}} & \makecell{\textbf{Write data} \\ \textbf{(ms)}} \\ \hline
Initial version & 584.65 & 14\% & 320.82 &  80.56 & 173.22\\ \hline
\makecell{Split out DRAM \\ connected ports} & 490.98 & 17\% & 256.76 &  80.56 & 140.65 \\ \hline
\makecell{Run concurrent loading and \\ storing via dataflow directive} & 189.64 & 30\% & 53.43  & 57.28 & 75.65\\ \hline
\makecell{Include X dimension of cube \\ in the dataflow region} & 522.34 & 10\% & 198.53 & 53.88 & 265.43\\ \hline
\makecell{Include X dimension of cube \\ in the dataflow region (optimised)} & 163.43 & 33\% & 45.65 & 53.88 & 59.86\\ \hline
256 bit DRAM connected ports & 65.41 & 82\% & 3.44 & 53.88 & 4.48\\ \hline
\makecell{256 bit DRAM connected ports \\ issue 4 doubles per cycle} & 63.49 & 85\% & 2.72 & 53.88 & 3.60\\ \hline
\end{tabular}
\caption{Profiling of HLS kernel and impact of DRAM based optimisations with a problem size of x=512,y=512,z=64 (16.7 million grid cells)}
\label{fig-kernel-dram-optimisation}
\end{figure*}

Timing for different versions of our kernel are summarised in Figure \ref{fig-kernel-dram-optimisation}, with the first entry, \emph{initial version}, being the HLS optimised kernel from  \cite{fpga} but with a slightly longer runtime due to modifications required for overlapping DMA transfer with compute as described in Section \ref{sec:DMA}. The integration of the profiler that was described in Section \ref{sec:prof} allowed us to break the overall timing down and explore the constituent contributions, validating our belief that the compute phase of the code was only responsible for a tiny fraction of the overall runtime. We found that only 14\% of the HLS kernel's runtime was actually spent in compute, and the rest of the time it was accessing DRAM, either reading or writing. Therefore, on average, for every 1 millisecond spent computing, the HLS kernel was spending over 7 milliseconds on memory access! 

It had been our hypothesis that moving from one 3D block to the next, as per Figure \ref{fig:domaindecomp}, would be the dominant factor, but this only accounted for 2.23 milliseconds and has been omitted from Figure \ref{fig-kernel-dram-optimisation} for brevity. Instead, the memory access required to read each individual slice from, and write each slices of results to, DRAM was dominant. Whilst this surprised us, on reflection maybe this is not so unexpected given that these slice based activities are performed much more frequently than moving from one 3D block to the next.

The code in Listing \ref{lst:hls_kernel} separated out reads to the \emph{U}, \emph{V}, and \emph{W} input variables, as well as writes to the corresponding \emph{SU}, \emph{SV}, and \emph{SW} output variables, into different loops. This is because all the variables shared a single HLS data port, and-so we modified the kernel to allocate a separate physical port for each variable. These were then tied together using an AXI crossbar at the block design level, and timing figures for this version are the second row of Figure \ref{fig-kernel-dram-optimisation}, \emph{split out DRAM connected ports}. This splitting out of the ports effectively allowed us to access DRAM via the variables in the same clock cycle, bringing the three reads into one single pipelined loop at the start of a slice, and the three writes into another single pipelined loop at the end of the slice. It can be seen that this reduced the data access time, but we were still only spending 17\% of the kernel execution time in compute, motivating a more fundamental restructuring of the code. 

\begin{figure}
\centering
\includegraphics[scale=0.57]{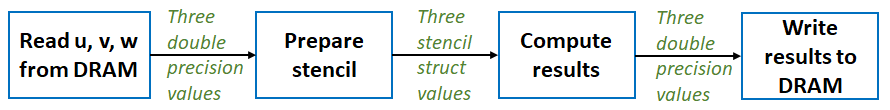}
\caption{Illustration of dataflow pipeline, streaming values between stages, each of which implements its own pipeline}
\label{fig:dataflow}
\end{figure}

Therefore we decided to redesign the kernel so that data could be read from, and written to, DRAM concurrently with the computation. The body of the loop over a 3D block (lines 9 to 46 of Listing \ref{lst:hls_kernel}) was extracted out into four separate functions. We structured these such that each function was performing a task independent from any other, apart from the fact that as the output of one task is generated this is then fed into the next task as its input. HLS's \emph{DATAFLOW} pragma was annotated to the function calls, and this indicates that the functions should be executed concurrently. The concurrent execution of each stage is illustrated in Figure \ref{fig:dataflow} and these stages are connected via HLS streams. A stream depth of 16 has been used, which means that the underlying FIFO queue implementation can buffer data if necessary, to somewhat decouple the stages from each other, for instance if one stage stalls. To send data to the next stage a \emph{write} call is issued on the appropriate HLS stream and this data is then consumed by a blocking \emph{read} call on the stream by that next stage. 

Because each stage is itself pipelined, this can be thought of as a pipeline of pipelines. The first stage in Figure \ref{fig:dataflow} reads individual double precision values for the \emph{U}, \emph{V}, and \emph{W} fields from DRAM and streams these to the next stage. Three separate streams, one for each field, are used as this means the three field values can be streamed to the next stage on the same clock cycle (as streams can only be written to once per cycle). Previously, the compute kernel itself retrieved the double precision 3D stencil data from BRAM required for computing a grid cell. Instead, the retrieval of stencil data and the actual compute has been split out into two separate stages. The second stage of Figure \ref{fig:dataflow}, \emph{prepare stencil}, builds up three stencil structures, one for each field grid cell, based upon data retrieved by the first stage and previous slices held in BRAM. These stencil structures are C \emph{structs} and as well as streaming this stencil data to the third, \emph{compute results}, stage, the second stage also updates the slice data held in BRAM to perform the shift from one slice to the next as it is iterating through the current slice's grid points. 


The third stage of Figure \ref{fig:dataflow} performs the actual computation based upon the stencil data it has received from the previous stage. For each grid cell this involves 53 double precision operations and the resulting double precision field values, \emph{SU}, \emph{SV}, and \emph{SW}, are streamed to the fourth stage as they are calculated, again using one stream per field. This fourth stage then writes the result values to DRAM. It can be seen from the third row of Figure \ref{fig-kernel-dram-optimisation} that this redesign had a very significant impact on overall performance, reducing the total runtime by over two and a half times. Not only did the DRAM memory access time reduce by over three times, the preparation of the stencil and compute time (the middle two stages) also reduced by almost a quarter. This reduction in the runtime of the middle stages surprised us and is because the splitting apart of the code has made it simpler, reducing the overall latency. 


Whilst this redesign of the code had reduced the runtime significantly, still only 30\% of the runtime was being spent in compute. The stages of Figure \ref{fig:dataflow}'s dataflow pipeline are only operating over a single slice, having to drain the entirety of the pipelines when moving from one slice in the 3D block to the next. Refactoring the code into dataflow stages had made it much simpler to then bring in the loop over the \emph{x} dimension, meaning that the pipelines will be filled for much longer. The runtime impact of this change is illustrated by the \emph{include X dimension of cube} row of Figure \ref{fig-kernel-dram-optimisation}. To our surprise, initially this modification very significantly increased the overall runtime, and specifically the time to load data from and write data to DRAM. 

The problem was that memory access on a slice by slice basis, in the \emph{z} and \emph{y} dimensions, is contiguous. As such, previously HSL issued a single read request, \emph{readreq}, for each of the three fields on its AXI-4 port. Then, within the pipelined loop over the slice only individual \emph{read} calls were issued. This was very important because the \emph{readreq} takes 25 cycles, whereas a \emph{read} takes 1 cycle. However, when we included the \emph{x} dimension the data access was no longer contiguous (see Figure \ref{fig:domaindecomp}) and as-such the \emph{readreq} was moved within the pipeline for every individual DRAM memory access, increasing the latency from 3 cycles to 28 cycles. The situation was worse for writing results to DRAM as that operation also requires the issuing of a write response, in that case increasing the latency from 3 cycles to 37 cycles. The fix was simple, we split apart each function of the first and fourth stages into two, the outer function looping over the \emph{x} dimension and calling into an inner function which loops over each slice in the \emph{z} and \emph{y} contiguous dimensions. By doing this, HLS once again recognised that data access in a single \emph{z} by \emph{y} slice is contiguous and placed the calls most effectively. This is the fifth row of Figure \ref{fig-kernel-dram-optimisation} and this has improved the runtime of all major parts of the code.

The DRAM controllers in the block design are 256 bits wide, whereas the HLS kernel was generating double precision ports of width 64 bits. Not only did the connection between the HLS kernel and DRAM need to pass through an AXI width converter at the block design level, adding at-least a cycle of latency, but we were also throwing away bandwidth as 256 bit is the optimum access width for the board's memory banks. Therefore we decided to make the HLS port width 256 bits, effectively reading or writing 4 double precision values per access. To do this we defined a C \emph{struct} containing an array of four double precision numbers. Then the type of the HLS array variables was changed from \emph{double} to this \emph{struct} and the HLS \emph{DATA\_PACK} pragma was used to pack the structure into a single scalar of width of 256 bits. Integrating this into the code required modifications to the first and fourth stages of Figure \ref{fig:dataflow}'s dataflow pipeline, packing and unpacking the structure. This provided a very significant reduction in memory access time and is illustrated by the \emph{256 bit DRAM connected ports} row of Figure \ref{fig-kernel-dram-optimisation}. 

\begin{figure}
\centering
\includegraphics[scale=0.6]{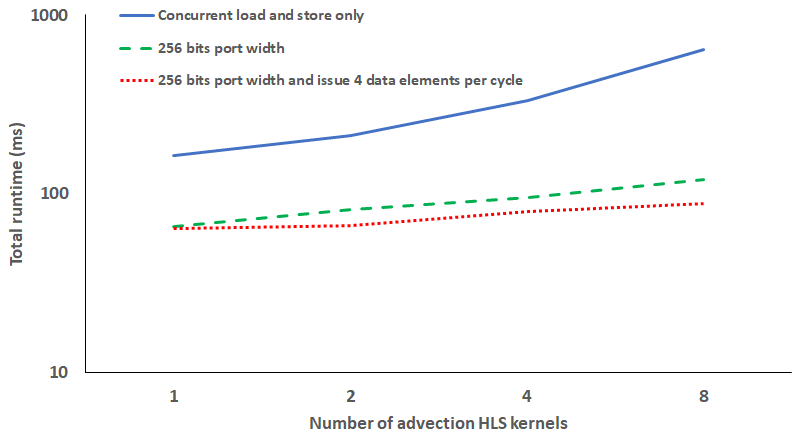}
\caption{Aggregate HLS kernel runtime as the number of HLS advection kernels is increased. For a problem size of x=512,y=512,z=64 (16.7 million grid cells)}
\label{fig:kernelscaling}
\end{figure}

The streams connecting the stages of our dataflow pipeline can not be written to, or read from, more than once in a single clock cycle. As such it was not possible to read the 256 bit structure and then write the four double precision values for a field to the stream in the next cycle. Instead, only one value could be written per cycle. Whilst the second stage of the dataflow pipeline can only consume, and the third stage only generate, one value for each field per cycle, being able to issue four double precision numbers for each field per cycle from the first stage, and similarly the fourth stage consume four values per field per cycle, means that the next DRAM memory access can occur sooner. As such we replicated the streams connecting the first stage to the second stage, and the third stage to the fourth stage, by four. The middle two stages simply cycle round the four streams for each field and whilst this had a marginal impact on the performance of a single HLS kernel, it was much more important when we ran multiple kernels concurrently. This is illustrated in Figure \ref{fig:kernelscaling} (log scale) which depicts the aggregate runtime of all HLS kernels based on a problem size of 16.7 million grid cells, as the number of kernels is increased (it does not include DMA transfer time). The approach of concurrently reading and writing, with the \emph{x} dimension in the dataflow pipeline, but without 256 bits width ports, resulted in an aggregate runtime that scales very poorly as the number of kernels is increased. This is due to contention on the DRAM and the adoption of 256 bits width, making use of full memory bandwidth, helps significantly but, the aggregate runtime with eight kernels is still almost double that of one kernel. Moving to four streams per field so that the first and fourth stages work with four doubles per field for each cycle helped significantly, the aggregate runtime with eight kernels only 20\% greater than one kernel. The reason for this is that each stream is of depth 16, so effectively for each field there can be 64 double precision numbers queued up and ready for the second stage to consume one cycle at a time. As-such, if there is contention and reads stall, then the buffer of existing data hides this to some extent. Furthermore, the fact that we are issuing 4 pieces of data per cycle into the buffer but only consuming it at rate of 1 piece of data per cycle means that if DRAM contention does occur, then the buffer fills back up quickly once contention has been reduced. In this manner, the impact of multiple kernels competing for DRAM access has been reduced.


The optimisations described in this section around memory accesses has meant that the runtime of our HLS kernel has reduced over nine times, with the compute utilisation going from 14\% to 85\%. Furthermore the restructuring of the code has meant that we have reduced the compute time itself by a quarter. These are significant performance improvements and are pivotal to the performance characteristics explored in Section \ref{sec:performance}.

\section{Ameliorating DMA data transfer}
\label{sec:DMA}
Previously our design involved performing all DMA data transfer to the PCIe board and then running the kernels only once this had completed. A similar situation was the case for the transfer of results off the board, where this would only occur once all kernels had completed their calculations. For larger problem sizes we found in \cite{fpga} that this was resulting in a very significant overhead, where DMA transfer time was responsible for over 70\% o the overall runtime. 

To address this, we split up the data on the host into chunks with the DMA transfer working on a chunk by chunk basis. This involves launching non-blocking DMA writes for each individual chunk on the host, and then tracking the completion status of the data transfer. As soon as the transfer of a chunk has completed, if there is an idle HLS advection kernel then this is started immediately, if no idle kernel is available then the chunk is queued for one to become free. Kernels are monitored, and as they complete a non-blocking DMA read is issued which copies back the result data to the host and the kernels are returned to a pool for re-use processing other chunks if necessary. Once all the kernels and DMA transfers copying back the result data have completed, then the MONC timestep can proceed. 

\begin{figure}
\centering
\includegraphics[scale=0.4]{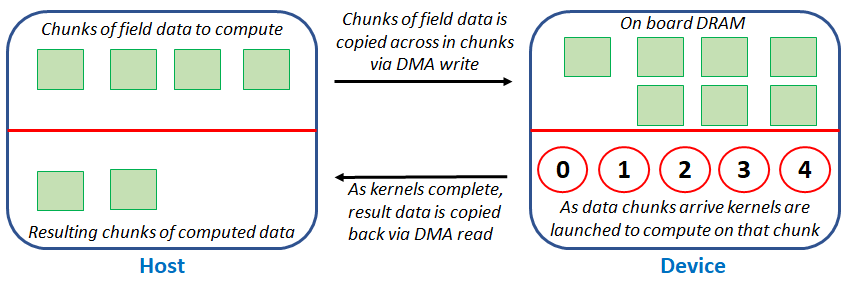}
\caption{Illustration of field data split up into separate chunks with idle kernels launched as the data arrives. Results are copied back onto the host as soon as the kernel has finished computing.}
\label{fig:datachunks}
\end{figure}

Effectively this is implementing an approach similar to the idea behind CUDA streams, where one could consider that there is an individual stream for each chunk of data, which involves transferring this chunk to the PCIe card, executing required computation on it, and then transferring the data back to the host. Figure \ref{fig:datachunks} illustrates this approach on the FPGA further, where the individual green boxes represent the chunks of data that the three fields of 3D atmospheric data has been split up into. These chunks are copied onto the PCIe card individually and kernels, 0 to 4 in this example, compute on chunks as they become available. Once a kernel has completed the host will initiate a non-blocking DMA transfer to copy results back. Whilst we can not entirely hide the cost of DMA transfer, for instance the first few input chunks and last few result chunks will need to be waited on regardless, the vast majority of DMA data transfer will occur whilst the kernels are busy computing, thus overlapping data transfer with compute.

It should be noted that, based on the API calls issued from the host, the pages of memory to be transferred are pinned and as such will never be swapped out to disk. Whilst Alpha Data's API also supports DMA transfers with non-pinned memory, the pinned calls provide significantly better performance. This is because DMA engines can start reading from, or writing to, the appropriate physical main memory locations straightaway, rather than the host having to ensure all pages are swapped in from disk before the engines  can begin. 

\section{Performance comparison}
\label{sec:performance}
\begin{figure}
\centering
\includegraphics[scale=0.4]{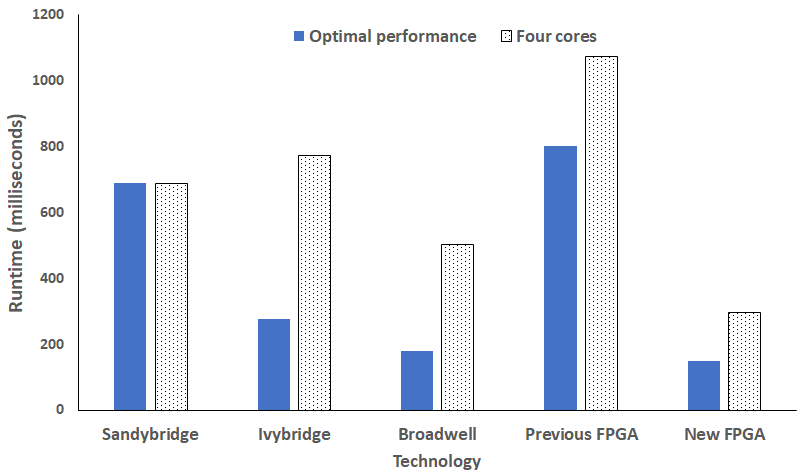}
\caption{Performance comparison of x=1024, y=1024, z=64 (67 million grid points) with a standard status cloud test-case}
\label{fig:performance-overall}
\end{figure}

We have investigated the performance of our approach using a standard MONC stratus cloud test case with a grid size x=1012, y=1024, z=64 (67 million grid cells). Figure \ref{fig:performance-overall} illustrates this performance comparison where we compared the performance of our new FPGA advection design against the previous FPGA design in \cite{fpga} and a C version of the same PW advection algorithm, threaded via OpenMP across the cores of the CPU (Sandybridge, Ivybridge, and Broadwell). For all runs the host code was compiled with GCC version 4.8 at optimisation level 3 and the results reported are averaged across fifty timesteps. For each technology there are two runtime numbers reported in milliseconds. The first, \emph{optimal performance}, illustrates the best performance we can get by threading over all the physical CPU cores (4 in the case of Sandybridge, 12 in the case of Ivybridge, 18 in the case of Broadwell) or the advection kernels (12 in the case of our previous FPGA design and 8 in the case of our new design.) We also report a four core number, which includes only running over four physical cores, or PW advection kernels in the case of the FPGA designs, as this is the limit of the Sandybridge CPU and allows a more direct comparison. 

The reason we leveraged twelve advection HLS kernels previously and only eight now is that the new HLS kernel design described in Section \ref{sec:optimisinghls} significantly increases the LUT and BRAM usage, meaning that we are now only able to fit eight kernels on the FPGA. However, the very significant increase in HLS kernel performance of over 9 times, along with overlap of DMA transfer, means that overall the performance of our new approach with eight kernels is over four times that of the previous approach with twelve kernels. Unlike the previous FPGA approach, when limited to four cores or kernels our new FPGA approach significantly out performs all other technologies. Four of our advection kernels are almost twice as fast as four cores of Broadwell.

With the optimal performance experiment, our HLS kernels are outperforming 18 cores of Broadwell (148 ms against 180 ms), and our new FPGA approach also out performs the other two CPU technologies at this configuration. Eight HLS kernels are outperforming eighteen cores, and whilst it might seem that if we could fit more kernels onto the FPGA then performance would be even higher, it should be noted that the overhead of DMA transfer accounts for 42\% of FPGA runtime at this problem size. This is much lower than the previous FPGA approach in \cite{fpga}, where the cost of DMA transfer was over 71\% of the runtime, but having sped up the HLS kernel runtime so considerably means that waiting for the transfer of the first few chunks of data before kernels start and the last few after all complete, is still a significant fraction of the overall runtime.

The CPU C code is directly based on the highly optimised Fortran code run in production and vectorised using the OpenMP \emph{simd} directive. It might seem strange that we ported the Fortran code into C for comparison, but we did this because the FPGA versions are driven by C code called from the model using ISO C bindings. Hence we felt that using the same programming language for the CPU comparison was the fairest experiment, and negated any differences between C and Fortran, or overheads of ISO C bindings. that are not the main focus of this work. Nevertheless, it should be noted that we found a negligible performance difference between the C and Fortran CPU versions of this kernel. 

\begin{figure}
\centering
\includegraphics[scale=0.45]{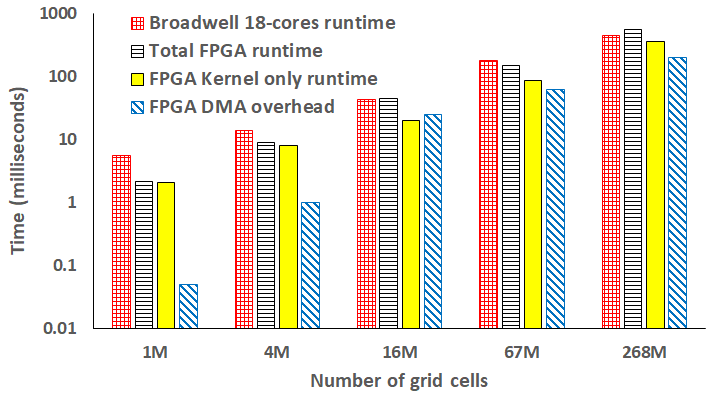}
\caption{Runtime of FPGA PW advection code (8 kernels) vs 18 cores of Broadwell against grid size with a standard stratus cloud test-case. For our FPGA approach we report three numbers, the total FPGA runtime, the execution time of the kernel alone (FPGA kernel only runtime) and the DMA transfer overhead time (FPGA DMA overhead)}
\label{fig:scaling-grid}
\end{figure}

Figure \ref{fig:scaling-grid} illustrates how the time, in milliseconds, changes as we scale the number of grid cells. For our FPGA approach (8 kernels) we report three numbers, the total FPGA runtime, the execution time of the kernel alone (\emph{FPGA kernel only runtime}) and the DMA transfer overhead time (\emph{FPGA DMA overhead}). We compare against 18 cores of Broadwell and for smaller grid sizes of 1 and 4 million grid cells our approach is 2.59 and 1.52 times faster than the CPU respectively. The two approaches are comparable at 16 million grid points and our approach again outperforms the Broadwell by 1.22 times at 67 million grid points. However, Broadwell out performs our FPGA approach by 1.23 times at 268 million grid points.

It should be noted that at all grid sizes, the FPGA kernel execution time alone is significantly smaller than the execution time of 18 Broadwell cores. However, as we increase the problem size the waiting for critical chunks of data to be transferred is a source of over 40\% overhead at 268 million grid points, whereas at a grid size of 1 million points it only accounts for 2\% of the total runtime. Figure \ref{fig-flops} illustrates the performance in GFLOP/s for our FPGA HLS kernels alone, the entire FPGA approach including DMA transfer, and 18 cores of Broadwell. For context, the previous FPGA version in \cite{fpga}, at 268 million grid points, achieved a total of 4.2 GFLOP/s and 14.4 GFLOP/s for the kernel alone. It can be seen that our HLS kernel significantly out-performs 18 cores of Broadwell, and the inclusion of DMA transfer overhead is limiting the performance at larger grid sizes. It is also interesting that the FPGA approach provides a much more consistent GFLOP/s performance level across all problem sizes than the Broadwell CPU. Based upon on-board sensors, the configured but idle total power draw of the ADM-PCIe-8k5 board is 28.9 Watts and this increases to 35.7 Watts under full load with the largest problem size when our advection kernels are running. Vivado estimates that the power draw of the design is 23 Watts. Whilst we did not have measuring equipment fitted to the CPUs, for reference the TDP of the 18 core Broadwell is 120 Watts \cite{intel_productsheet}.

\begin{figure}
\small
\begin{center}
\begin{tabular}{ | c c c c | }
\hline
\textbf{Grid size} & \makecell{\textbf{FPGA Kernel} \\ \textbf{GFLOP/s}} & \makecell{\textbf{Total FPGA} \\ \textbf{GFLOP/s}} & \makecell{\textbf{Broadwell} \\ \textbf{GFLOP/s}} \\ \hline
1M & 25.2 & 24.7 & 9.5\\
4M & 26.5 &  23.6 & 15.4\\
16M & 42.4 & 18.8 & 19.6\\
67M & 39.4 & 22.9 & 18.8\\
268M & 38.1 & 24.4 & 30.2\\
\hline
\end{tabular}
\end{center}
\caption{GFLOP/s performance of FPGA advection scheme (8 kernels) against 18 cores of Broadwell}
\label{fig-flops}
\end{figure}

\section{Conclusions and further work}
\label{sec:conc}
In this paper we have described our approach in optimising an existing FPGA port of an atmospheric model's advection scheme. We focused around the overhead of data transfer and this involved two main areas, reducing the runtime of the HLS kernel and ameliorating the cost of DMA transfer to and from the PCIe card. In order to optimise the HLS kernel we first had to understand where the bottlenecks lay and as such profiled the HLS code, block by block. We explored how this information was used to reduce the runtime, exploring the impacted on time spent in different parts of the kernel. Our strategy for improving the DMA transfer time was to chunk the data up and overlapping transfer with compute where possible.

These optimisations have meant that our HLS kernel alone runs over nine times faster than the version developed in \cite{fpga}. More generally, the MONC advection scheme on the FPGA is now competitive against popular CPUs, out-performing the 18-core Broadwell for all but the largest problem size. Whilst the Broadwell does still out-perform our FPGA approach at the largest grid size of 268 million cells, over 40\% of the FPGA runtime was in DMA transfer overhead. Bearing in mind that a total of 12.88GB of data must be transferred in this case, an overhead of 200 milliseconds is not so unreasonable.

It is this DMA transfer overhead that we are looking to tune as part of further work, queuing up chunks of data to kernels, rather than waiting on the host for kernels to become free. Building further on the idea of CUDA streams, we think that by driving more from the FPGA, with an IP block controlling kernel activation rather than the host side, further performance improvements should be possible. Another area of interest is the use of larger FPGAs, which will allow us to leverage more advection kernels and smaller data chunk sizes, as well as moving from double to single precision. This last point is important because, now that the compute kernel time takes 85\% of the kernel's runtime, changing the precision will have a much greater overall impact than it would have when the compute time only accounted for 14\% previously.

We conclude that the use of FPGAs is very exciting and competitive with CPUs. Whilst there is undoubtedly still room for some further improvement, the data movement optimisation techniques described in the paper have illustrated the major architectural changes needed at a code level to bring the performance to a practical level for a real use and exploit the reconfigurable computing architecture provided by FPGAs. The HLS kernel we have ended up with runs over 800 times faster than the initial version in \cite{fpga} which was directly based on the CPU code with no modifications or optimisations. This really illustrates how programmers must \emph{think dataflow} for FPGAs and not only should the entire algorithm, including the fetching of data, be pipelined, but also extra attention must be given to physical aspects, such as the width of memory interface and location of data to be used. 

\section{Acknowledgements}
The authors would like to thank Alpha Data for the donation of the ADM-PCIe-8K5 card used throughout the experiments of work. This work was funded under the EU FET EXCELLERAT CoE, grant agreement number 823691.


\bibliographystyle{./bibliography/IEEEtran}
\bibliography{./bibliography/IEEEexample}

\appendices

\section{Artifact Description Appendix}

\subsection{Description}

\subsubsection{Check-list (artifact meta information)}

{\small
\begin{itemize}  
  \item {\bf Program: } C and Fortran
  \item {\bf Compilation: }GCC version 4.8 with -O3, Vivado HLS version 2018.3
  \item {\bf Data set: }Standard MONC status test-case configuration file
  \item {\bf Run-time environment: } A variety of machines were used for comparison, all running Linux, MPICH and OpenMP. For the FPGA code Alpha Data'a ADMXRC3 API version 1.8.2 was used.
  \item {\bf Hardware: } Eight-core Sandybridge based system used to host the FPGA, four cores of this were used for the CPU only Sandybridge experiment (4 cores per NUMA domain.) Twelve Ivy Bridge cores of a Cray XC30 was used for the Ivy Bridge experiment, and eighteen Broadwell cores of an HPE/SGI 8600 system. An ADM-PCIe-8K5 board was used which mounts the Kintex Ultrascale KU115-2 FPGA.
  \item {\bf Binary: } MONC CPU versions require OpenMP, for the FPGA version this requires AlphaData's API library and board support. Xilinx tooling requires to synthesise the HLS kernel and generate the bitstream from Vivado.
  \item {\bf Execution: } We built and executed the executable from Linux.
  \item {\bf Output: } From the MONC test-case timings are produced and an output file of field data is generated which can be used to check for consistency
  \item {\bf Publicly available?: }No
\end{itemize}
}

\subsubsection{Hardware dependencies}
Any machine running Linux with appropriate FPGA PCIe card installed
\subsubsection{Software dependencies}
GCC version 4.8, the support libraries installed for the board and ADMXRC3 API version 1.8.2 to interact with the board. In this case these came from Alpha Data as the ADM-PCIe-8K5 is one of their products. Xilinx tooling (HLS and Vivado) required along with appropriate licences to generate the bitstream.
\subsubsection{Datasets}
For the experiments here we used the standard MONC stratus cloud test case.
\subsection{Installation}
We synthesised our kernel using HLS and then exported the bitstream. Next the block design was build as appropriate for the specific FPGA technology, and we re-used the shell for different HLS kernels, as such it was often just a case of simply importing the updated IP block and Vivado refreshing the kernels in the block design. The bitstream was then generated and flashed to the PCIe card, this required a restart of the card sometimes involved restarting the entire machine. On the host side, we used the addresses generated by HLS to set the appropriate values and transfer the field data to card DRAM via Alpha Data's DMA transfer non-blocking calls as part of their API, before starting the kernels as chunks arrive.

\subsection{Experiment workflow}
\begin{enumerate}
\item Develop the appropriate HLS kernel 
\item Use Vivado HLS to synthesise this and export as IP block
\item Import into the block design (the shell) and make any modifications required if the experiment demands it
\item Generate the bitstream
\item On the host side ensure the MONC code is interacting with the appropriate addresses as reported by HLS.
\end{enumerate}

\subsection{Evaluation and expected result}
We compared our results against a threaded version of the advection scheme running on the cores of the different CPU technologies. Whilst the MONC code is in Fortran, we ported the advection kernel to C and use that version for comparison because the reconfigurable version uses C to interact with the FPGA and-so we felt that this was the fairest experiment. All results have been checked at the grid point level to ensure that they are producing consistent results between the different versions of the code and that they are actually calculating the same quantities. 

\subsection{Experiment customization}
It is, of course, possible to experiment with the HLS kernels and use these to run larger system sizes if you have an FPGA board with more on-card DRAM than the 16GB provided by the ADM-PCIe-8K5. It is also possible to run with an increased number of advection kernels if you have a larger FPGA than the Kintex Ultrascale KU115-2, or modification of the kernels for the reader's own problems can also be made.

\section{Artifact Evaluation}

\subsection{Results Analysis Discussion}
On the CPU we use the \emph{gettimeofday} C call which provides microsecond resolution timings. For FPGA runs these timings correlate closely with those that we get from our profiling IP block which is clock cycle accurate. All results were checked, grid point by grid point, for consistency between the FPGA and CPU versions to ensure that they are calculating the same quantities to ensure a fair experiment. Runtimes were averaged over 50 timesteps and power consumption figures were generated from sensors providing voltage and amperage for the 12V and 3.3V power rails on the board. There is more work to be done understanding the power usage of the ADM-PCIe-8K5 board and specifically what the exact overhead of the FPGA execution is when the kernels are running. To be explicit we have quoted the entire power usage of the PCIe board when configured but idle and when the kernels are active. 

\end{document}